\documentclass[%
 reprint,
 superscriptaddress,
 amsmath,amssymb,
 aps,
]{revtex4-2}

\usepackage{graphicx}
\usepackage[caption=false]{subfig}
\usepackage{dcolumn}
\usepackage{bm}
\usepackage[colorlinks=true,linkcolor=blue,urlcolor=black,citecolor=blue,bookmarksopen=true]{hyperref}
\usepackage{amsmath,physics}
\usepackage{xcolor}
\newcommand{\mol}{\text{mol}}

\begin{document}


\title{Minimizing state preparation times in pulse-level variational molecular simulations}

\author{Ayush Asthana}
 \affiliation{Department of Chemistry, Virginia Tech, Blacksburg, VA 24061, USA}
\author{Chenxu Liu}%
\affiliation{Department of Physics, Virginia Tech, Blacksburg, VA 24061, USA}
\author{Oinam Romesh Meitei}
\affiliation{Department of Chemistry, Massachusetts Institute of Technology, Cambridge, MA 02139, USA}
\author{Sophia E. Economou}
\affiliation{Department of Physics, Virginia Tech, Blacksburg, VA 24061, USA}
\author{Edwin Barnes}
\affiliation{Department of Physics, Virginia Tech, Blacksburg, VA 24061, USA}
\author{Nicholas J. Mayhall}
 \email{nmayhall@vt.edu}
\affiliation{Department of Chemistry, Virginia Tech, Blacksburg, VA 24061, USA}





\date{\today}

\begin{abstract}
Quantum simulation on NISQ devices is severely limited by short coherence times. A variational pulse-shaping algorithm known as ctrl-VQE was recently proposed to address this issue by eliminating the need for parameterized quantum circuits, which lead to long state preparation times. Here, we find the shortest possible pulses for ctrl-VQE to prepare target molecular wavefunctions for a given device Hamiltonian describing coupled transmon qubits. We find that the time-optimal pulses that do this have a bang-bang form consistent with Pontryagin’s maximum principle. We further investigate how the minimal state preparation time is impacted by truncating the transmons to two versus more levels. We find that leakage outside the computational subspace (something that is usually considered problematic) speeds up the state preparation, further reducing device coherence-time demands. This speedup is due to an enlarged solution space of target wavefunctions and to the appearance of additional channels connecting initial and target states.
\end{abstract}


\maketitle

\section{\label{sec:level1}Introduction}
 Noisy intermediate-scale quantum (NISQ) devices have qubits with low coherence times and frequent gate errors. This constrains today's quantum algorithms to short circuit depths and has inspired many algorithms aimed at minimizing the quantum resources needed  to obtain accurate results~\cite{kandala2017hardware,huggins2020non,grimsley2019adaptive,tang2021qubit,lee2018generalized,ryabinkin2018qubit,gard2020efficient}.
 Despite rapid advances on both the algorithmic and hardware fronts, the path to useful quantum advantage remains elusive \cite{arute2019quantum,zhong2020quantum,maslov2021quantum,nam2019low,cerezo2021variational,cao2019quantum,tilly2021variational,fedorov2022vqe}.
Thus, generating unitary transformations in a short time on a quantum device is of fundamental and practical importance in quantum computing.

Because most qubit platforms contain at least two levels, 
one must typically designate two of the states (i.e., $\ket{0}$ and $\ket{1}$) as the ``computational basis''
(the states to which the computational problem is mapped). 
Any uncontrolled population of states outside of the computational space (referred to as ``leakage'') is generally harmful from a computational perspective.
For example, superconducting transmon qubits are one of the leading approaches for realizing scalable \cite{acin2018quantum} quantum devices with high fidelity and low noise in gates~\cite{kjaergaard2020superconducting}.
However, due to the typically low anharmonicity of the transmon qubits, higher energy levels are easily populated. 
While these higher energy states can sometimes be leveraged as an additional quantum resource~
\cite{neeley2009emulation,lanyon2009simplifying,bechmann2000quantum,cerf2002security,blok2021quantum,rosenblum2018fault,chow2013microwave,wu2020high,ashhab2021speed}, 
if the quantum logic gates can be extended to explicitly act on the increased computational space, 
leakage is generally harmful, leading to difficulties in error mitigation and measurement~
\cite{varbanov2020leakage,ghosh2013understanding,fowler2013coping}.
To minimize leakage, one typical approach is to constrain the power of the drive or to use optimal control techniques~\cite{motzoi2009simple,werninghaus2021leakage} to implement fast quantum gates with high fidelity on transmon qubits \cite{economou2015analytical,deng2017robustness}.

On the algorithms side, variational quantum algorithms (VQAs) \cite{peruzzo2014variational} are among the most promising approaches for solving a range of both optimization and simulation problems on NISQ era devices due to their resilience to noise \cite{sharma2020noise}.
Although VQAs have been developed for applications including quantum chemistry \cite{mcardle2020quantum,cerezo2021variational,tilly2021variational}, optimization problems \cite{farhi2014quantum,farhi2016quantum}, quantum neural networks \cite{farhi2018classification,hou2021universal,schuld2014quest,jeswal2019recent} and many more, our focus is on the electronic structure problem, which deals with finding the low-lying energy states of a molecule~
\cite{mcardle2020quantum,cerezo2021variational,cerezo2021variational}.
The electronic structure problem is the cornerstone of quantum chemical calculations, which are believed to be one of the leading fields in which quantum computers can demonstrate an advantage over classical computers. 
Molecular simulations on a quantum computer have already reproduced some known results for small molecules using variational quantum eigensolver (VQE) algorithms \cite{kandala2017hardware}. 
VQEs, a subset of VQAs, are classical-quantum algorithms that approximate the ground state of a many-body Hamiltonian by optimizing a parameterized quantum circuit that prepares a trial state wavefunction ($\ket{\Psi(\theta)}$) to minimize the molecular energy. The optimal parameter values are given by
\begin{align}
    \theta^*=\text{arg min}_{\theta}(\langle \Psi(\theta)|\hat{H}_{\text{mol}}|\Psi(\theta)\rangle), 
\end{align}
where $\hat{H}_{\text{mol}}$ is the molecular Hamiltonian. 

It has recently been recognized that VQAs can be both better understood and significantly improved by establishing connections with quantum optimal control theory (QOC) \cite{Meitei2020,magann2021pulses,yang2017optimizing,choquette2021quantum,larocca2021diagnosing}. 
VQAs involve the optimization of parameters in discrete quantum circuit logic elements, whereas QOC functions more broadly to optimize a time-dependent drive Hamiltonian, typically by determining optimal pulse shapes of applied fields. 
Every parameterized quantum circuit corresponds to a pulse, but not all pulses have an associated parameterized quantum circuit. 
Although the two approaches share the same objective, to reach the final state that minimizes a cost function (e.g., molecular energy), 
QOC-based VQAs are inherently more powerful than circuit-based VQAs.


This connection is directly exploited by the recently developed ctrl-VQE algorithm~\cite{Meitei2020}, which defines a QOC-based ansatz at the device level, taking pulse parameters as optimization parameters and performing VQA optimization steps to minimize the molecular energy. 
Ctrl-VQE has been shown to reduce the coherence-time requirements for molecular simulations by several orders of magnitude, therefore showcasing the potential to simulate strongly correlated systems on NISQ-era quantum computers. 

In addition to the reduction in coherence-time demands demonstrated in Ref.~\cite{Meitei2020}, connections between QOC and VQAs also provide critical conceptual insights regarding the complexity of numerical optimization.
For controllable quantum systems with unconstrained quantum resources, the control landscape has been shown to be free from local minima \cite{russell2017control}. The absence of local minima is important for the optimization process to reach the target state without getting stuck in traps. On the other hand, such traps often occur in parameterized gate-based VQAs unless more than a critical number of parameters are included (referred to as overparameterization) \cite{larocca2021theory}. A conceptual understanding of numerical optimization is important for the development of scalable quantum computing algorithms.

The minimum evolution time (MET) required to rotate one quantum state to another has been studied using both constrained and unconstrained controls \cite{deffner2017quantum,poggi2020analysis,tibbetts2012exploring}. 
The field required to drive a MET transformation forms the ``time-optimal'' control, and the parameters for this field are the time-optimal control parameters. 
Coming from classical control theory, Pontryagin's maximum principle has been applied to study optimal control for VQAs with a continuous-time evolution, using an amplitude-constrained drive \cite{yang2017optimizing, brady2021optimal,lin2019application}.
Yang et al.~\cite{yang2017optimizing} found that the time-optimal controls exhibit a bang-bang form (control parameters saturate the bounds) due to the linear dependence on control waveforms in the drive Hamiltonian. Brady et al.~\cite{brady2021optimal} showed that more general optimal waveform shapes may exist (depending on the problem), as a key assumption in earlier work is not valid universally. 

In this work, we show through numerical simulations that ctrl-VQE yields time-optimal control fields that prepare target molecular ground states in the shortest time possible for a given transmon device. These optimal pulses have a bang-bang form that is fully consistent with the Pontryagin maximum principle, as we confirm with analytical calculations. Using these time-optimal solutions, we investigate the effect of additional states outside the computational subspace, as naturally occur in transmon systems. We find that, surprisingly and contrary to typical scenarios, 
leakage outside of the computational space can significantly improve the performance of ctrl-VQE by further reducing the state preparation time. We show that this speedup can be attributed to two factors: (i) an expansion of the solution space of target wavefunctions, and (ii) the appearance of shorter quantum trajectories that are made accessible via the leakage states. These results constitute an important step toward overcoming coherence-time limitations of quantum simulation on NISQ devices.

This paper is organized in the following way. We first discuss the ctrl-VQE algorithm (Sec.~\ref{s1}) and time-optimal controls (Sec.~\ref{s12}). We then provide the computational details for the simulations in this paper in Sec.~\ref{s2}. Results related to time-optimal controls in ctrl-VQE (\ref{s31}) and faster state evolution using leakage states (\ref{s32}) are discussed in Sec.~\ref{s3}. Finally, we present the summary and outlook in Sec.~\ref{s4}.

\section{Algorithm and time-optimal controls}
\subsection{ctrl-VQE}\label{s1}
The ctrl-VQE approach uses a QOC-based wavefunction ansatz at the device level, instead of parameterized quantum circuits, for VQE simulations \cite{Meitei2020}. 
Because ctrl-VQE creates a device-specific ansatz built from parameterized control fields, the Hamiltonian of the device needs to be defined upfront. In our simulations, we use superconducting transmon qubits with constant, always-on, interqubit couplings.  The device Hamiltonian is given by
\begin{align}
    \hat{H}_{D}=\sum_{q}\omega_q a^{\dagger}_q a_q -\sum_{q}\frac{\delta_q}{2} a^{\dagger}_q a^{\dagger}_q a_q a_q +\sum_{\expval{pq}}g_{pq}a^{\dagger}_p a_q,
\end{align}
where $\omega_q$ is the transition frequency of the $\ket{0}\rightarrow \ket{1}$ transition (resonance frequency) for qubit $q$, $\delta_q$ is the qubit anharmonicity, $g_{pq}$ is the constant coupling rate for a pair of transmons, and the brackets in $\expval{pq}$ indicate a restricted sum over directly connected transmons. The control Hamiltonian that drives the system under the rotating wave approximation is given by
\begin{align}\label{eq0}
    \hat{H}_{C} (t)=\sum_{q}\Omega_q(t)(e^{i\nu_qt} a_q+e^{-i\nu_qt} a^{\dagger}_q),
\end{align}
where $\Omega_q(t)$ and $\nu_q$ are the time-dependent amplitude and (time-independent) frequency of the drive
on qubit $q$. The state evolves according to the time-dependent Schr\"{o}dinger equation in the interaction picture,
\begin{align}\label{eq1}
    \frac{\partial}{\partial t}\ket{\Psi_{I}(t)}=-i\hat{H}_{I,C}(t)\ket{\Psi_{I}(t)},
\end{align}
where the $\ket{\Psi_{I}(t)}$ is the wavefunction, and the
interaction picture control Hamiltonian is given by 
\begin{align}\label{eq2}
    \hat{H}_{I,C}(t)=e^{i\hat{H}_Dt}\hat{H}_{C}(t)e^{-i\hat{H}_Dt}.
\end{align}
To obtain the final state at the time, $T$, the time-dependent Schr\"{o}dinger equation is solved, while the optimization is carried out on the drive amplitudes and frequency ($\Omega(t)_n$, $\nu_n$ for each qubit), which are varied to minimize the molecular energy given by
\begin{align}\label{cost}
    E_{\mol}=\langle \tilde{\Psi}_I(T)|\hat{H}_{\mol}|\tilde{\Psi}_I(T)\rangle.
\end{align}
Note that $\hat{H}_{\mol}$ is the molecular Hamiltonian mapped to the qubit system, and $\ket{\tilde{\Psi}_I(T)}$ is the wavefunction at the final evolution time ($\ket{\Psi_I(T)}$) projected onto the computational subspace, which is the subspace spanned by the computational basis states ($\ket{0}$ and $\ket{1}$) of each qubit. The wavefunction in the computational basis is then normalized, although the unnormalized wavefunction can be used in the computations to eliminate any leakage. $T$ is the total time of evolution and is kept fixed during a simulation.
The ctrl-VQE ansatz can be summarized as
\begin{align}
    |\Psi_I^{\text{trial}}(\Omega_n(t), \nu_n)\rangle =\hat{\mathbb{T}} e^{-\int_{t=0}^{T}dt \hat{H}_{I,C}(t, \Omega(t)_n, \nu_n)}|\psi_{I}(0)\rangle, 
\end{align}
where $\hat{\mathbb{T}}$ is the time-ordering operator, and $\ket{\psi_{I}(0)}$ is the initial state, usually taken as the Hartree-Fock (HF) wavefunction. This trial wavefunction is evaluated at each iteration, and the optimization parameters are varied to minimize the molecular energy. In what follows, we suppress the subscript $I$ on the states for notational simplicity. 

\begin{figure}[t]
  \includegraphics[trim={1cm 0cm 7cm 0cm},clip,width=\linewidth]{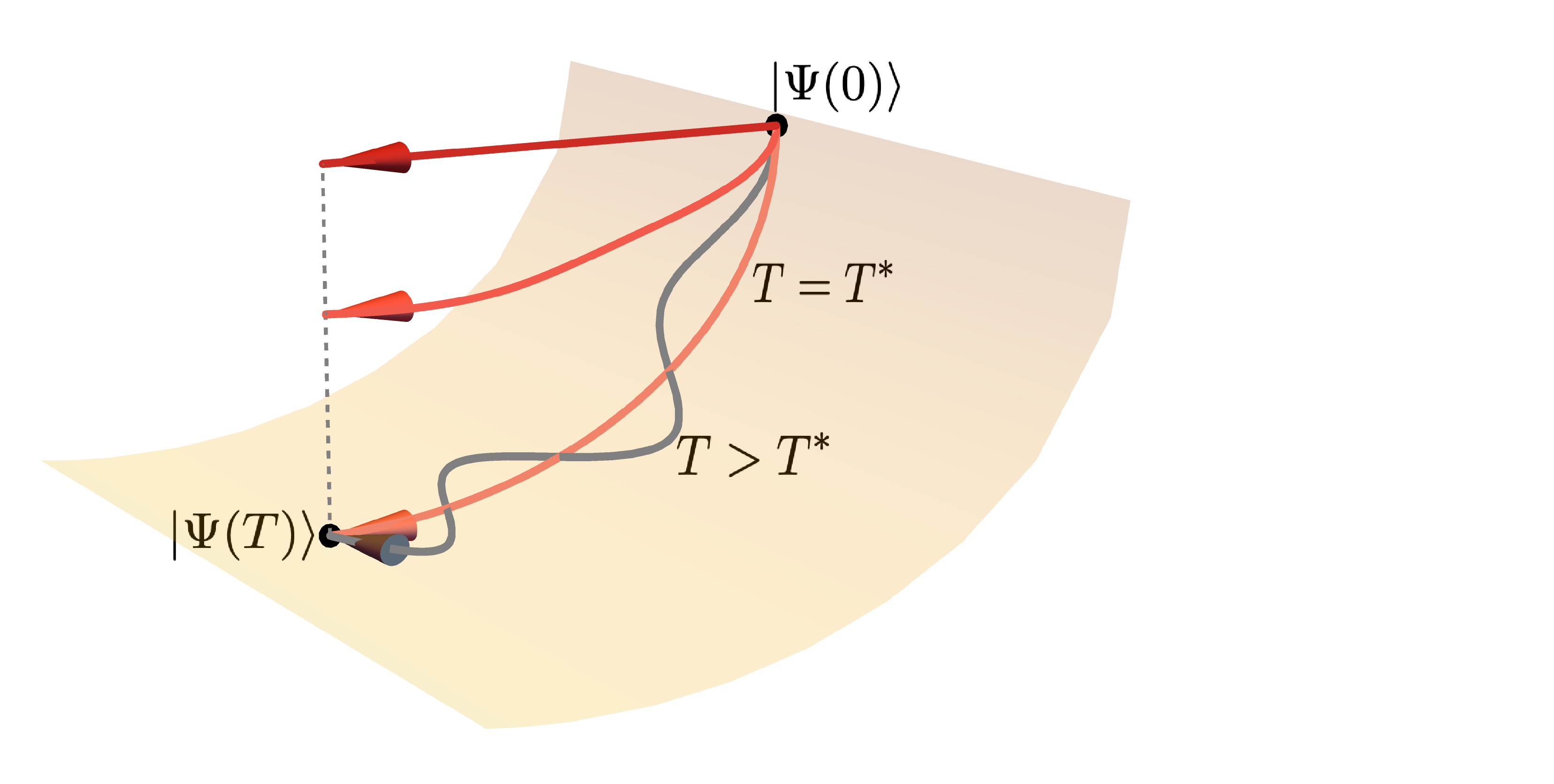}
    \caption{Qualitative depiction of state evolution in Hilbert space of qudits with two or more levels. The time-optimal path in the computational subspace (yellow surface) is shown as the path with evolution time $T=T^*$. Many paths are possible to reach the solution $|\Psi(T)\rangle$ when $T>T^*$ (gray curvy line). Access to leakage states provides new paths (shown by darker red paths) that reach the solution faster than $T^*$ when the final state is projected onto the computational subspace.}
    \label{fig1}
\end{figure}

\subsection{Minimum evolution time and time-optimal controls}\label{s12}

Control problems aim to find a target state that minimizes the cost function, which in the case of molecular simulation is the molecular energy [Eq.~\eqref{cost}]. In the qualitative depiction in Fig.~\ref{fig1}, this state is represented by $\ket{\Psi(T)}$. Starting from the initial state, represented by $\ket{\Psi(0)}$ (see Fig.~\ref{fig1}), the state evolves in time through the action of driving pulses until it reaches the target state. There exists a minimum pulse duration (the MET) that is required to carry out this state evolution. Control pulses with durations that exceed the MET ($T>T^*$ in Fig.~\ref{fig1} where $T^*$ is the MET for qubits) can drive the system along one of the infinitely many possible trajectories in Hilbert space to reach the target state, while time-optimal pulses drive the initial state to the target state along special trajectories in a time equal to the MET. 
The MET generally depends on the constraints imposed on the driving pulses. 
Ctrl-VQE generates controllable evolutions in the control theory sense, as the entire Hilbert space can be spanned to arbitrary precision by varying unconstrained parameters of the driving pulses. In ctrl-VQE, as the total pulse duration ($T$) increases, the Hilbert space accessible to the qubit system increases as well. 

In the present study, we numerically find $T^*$ and the associated time-optimal controls by starting at a large enough pulse duration and reducing the pulse duration until we find the shortest pulse at which the system can reach the target state to the desired accuracy. 
The pulse is divided into many step-wise constant segments with practical bounds on amplitudes and frequencies. Dividing the pulse into many adjustable segments provides a large amount of freedom in shaping the pulse, essentially allowing the ctrl-VQE algorithm to choose any possible optimal pulse shape in the optimization process. Because the numerical optimization of time-optimal controls is quite challenging, it is difficult to ensure that one has converged to $T^*$. 
As a guarantee, in what follows we compare our numerical solutions to analytical results obtained from Pontryagin's theory to confirm that our numerically obtained METs are correct.

\section{Computational details}\label{s2}
The calculations are carried out using a local version of the ctrlq code \cite{Meitei2020,Meigit}. As a test system for our simulations, we have chosen the problem of finding the ground state energy of the H$_2$ molecule at a bond distance of $1.5~\text{\AA}$, which has relatively strong correlations. Molecular integrals are generated using PySCF \cite{pyscf}  and STO-3G basis set is used in this work. We have used the parity mapping with Z2 symmetry reduction to map molecular Hamiltonian on the qubit basis. The initial state is taken as the unentangled HF state for this problem, and the target state is the exact ground state (i.e., full configuration interaction (FCI)). 
The numerical optimization of pulse amplitudes and frequencies is carried out using the l-BFGS-b optimization algorithm, and the pulses are taken to be piece-wise constant with 100 segments (unless otherwise stated). The time-discretized amplitude, $\Omega(t)$, is given by
\begin{align}
\Omega_k(t)=\begin{cases}
c_1&\: 0\leq t\leq t_1\\
c_2&\: t_1\leq t\leq t_2\\
\dots\\
c_n&\: t_{n-1}\leq t\leq T
\end{cases}
\end{align}
for qubit $k$, where $c_1,c_2\dots c_n$ represent amplitude parameters for each pulse segment, and $T$ is the final time. 
The pulse amplitude is constrained between $\pm 20$~MHz unless otherwise mentioned, which is based on common values for control signals in transmon qubits~\cite{Ku2017}.
The carrier signal frequency $\nu_k$ given in Eq.~\eqref{eq0} is constrained such that its difference from the qubit frequency, $\nu_k-\omega_k$, is in the range $\pm 1$~GHz, where $\omega_k$ is the qubit frequency.
A convergence error threshold of $10^{-8}$ Hartree in molecular energy is used in the optimization process, which corresponds to close to unit fidelity with the target FCI wavefunction. 
Since we encounter many local minima when the pulse duration is close to its minimum value, we consider MET to be the minimum pulse duration at which a solution could be reached in 1000 random initializations. 
The results are minimally affected by increasing the number of pulse segments or Trotter steps (see SI for details).
The parameters in the device Hamiltonian, $\omega_q$, $\eta_q$ and $g_{pq}$, used in the simulations are given in Table~\ref{tab1}. 

\begin{table}[t]
    \centering
    \begin{tabular}{c | c c}
    \hline\hline
        & \multicolumn{2}{c}{Transmon}\\
    \hline
          &1&2  \\
         $\omega/(2\pi)$ & 4.8080 & 4.8333\\
         $\delta/(2\pi)$ & 0.3102 & 0.2916\\
    \hline
    & \multicolumn{2}{c}{$1 \rightarrow 2$}\\
    \hline
         $g/(2\pi)$ & \multicolumn{2}{c}{0.01831}\\
    \hline\hline
    \end{tabular}
    \caption{Device parameters used in Eq.~\eqref{eq0} for a two-transmon system. All values are in GHz.}
    \label{tab1}
\end{table}

\section{Results and discussion}\label{s3}

In the following subsections, we discuss our main results related to time-optimal controls in the ctrl-VQE algorithm applied to transmon qubits. We first discuss the time-optimal pulse shape for ctrl-VQE through the application of Pontryagin's maximum principle and numerical simulations. Next, we qualitatively analyze the near-MET evolution of the two-qudit state, with a particular focus on the role of leakage to higher transmon levels outside the computational subspace.  

\subsection{Time-optimal control in ctrl-VQE}\label{s31}

\begin{figure*}[tbp]
    \centering
    \subfloat[qubit system \label{fig2a}]{
        \includegraphics[width = 0.45\textwidth,height=10cm,keepaspectratio]{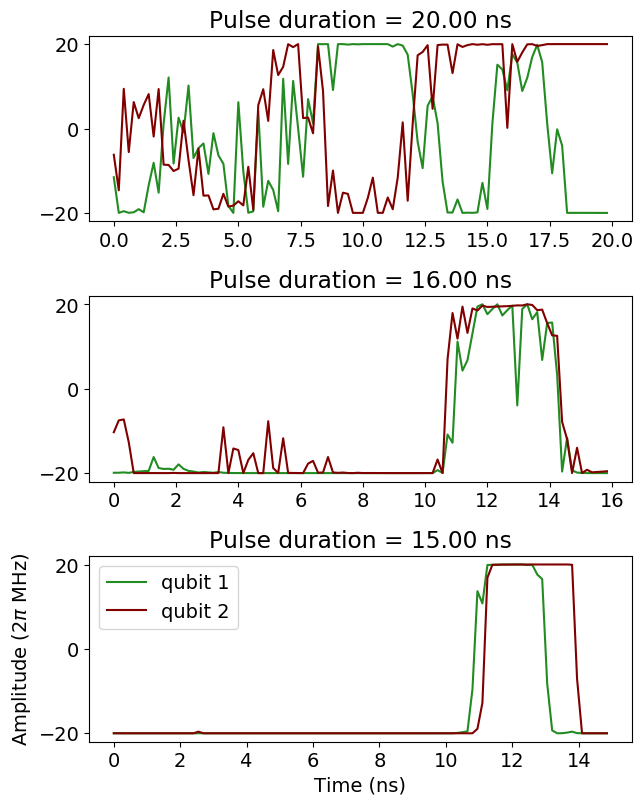}    
    }
    \subfloat[qutrit system \label{fig2b}]{
    \includegraphics[width = 0.45 \textwidth,height=10cm,keepaspectratio]{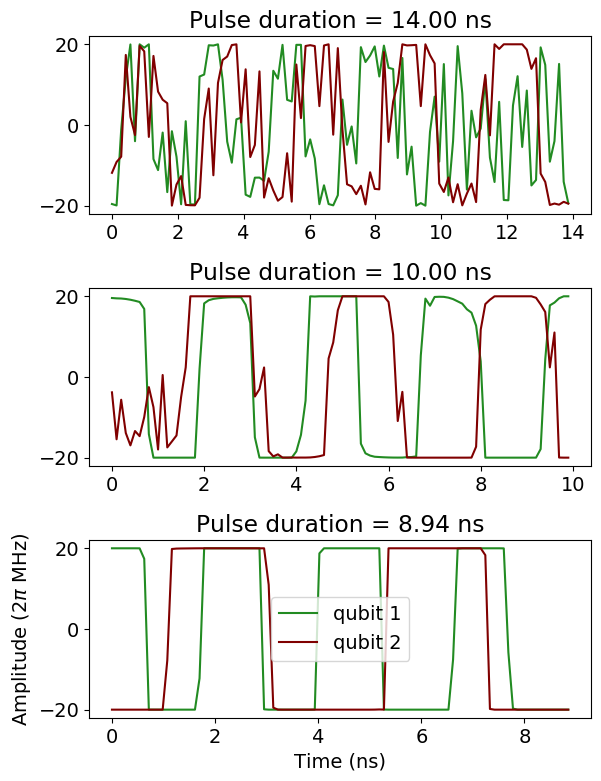}
    }
    \caption{(a) How the optimized pulse shape changes for a two-qubit system as the pulse duration is reduced from $20.00$~ns to $15.00$~ns. (b) How the optimized pulse shape changes for a two-qutrit system as the pulse duration is reduced from $14.00$~ns to $8.94$~ns. The H$_2$ ground state energy is calculated in these simulations at a bond distance of 
    $1.5~\text{\AA}$. 
    The pulse shape evolves into bang-bang form as the pulse duration is restricted towards the minimum evolution time (MET) needed to produce the ground state.}
    \label{fig2}
\end{figure*}

Time-optimal controls for quantum algorithms have been previously studied using quantum optimal control theory with continuous-time controls \cite{yang2017optimizing,brady2021optimal,lin2019application, Boscain2021}. 
Pontryagin's principle is applied to derive conditions under which an associated cost function is minimized at the MET. We will follow the same procedure and use Pontryagin's principle to find conditions for time-optimal controls for the ctrl-VQE algorithm. The dynamics of the multi-qudit state in ctrl-VQE is governed by the time-dependent Schr\"odinger equation given in Eqs.~\eqref{eq1}-\eqref{eq2}. 
The cost function in the case of the electronic structure problem takes the form of Eq.~\eqref{cost}. This can be formulated into a time-optimal control problem, with the modified cost function (not to be confused with the cost function associated with the ctrl-VQE algorithm) to be minimized as
\begin{align}
    J& = \langle \Psi(T)|\hat{H}_{\mol}|\Psi(T)\rangle \nonumber \\
    &+\int_0^{T}\langle \lambda(t)| \left( -i\hat{H}_{I,C}(t)|\Psi(t)\rangle-|\dot \Psi(t)\rangle \right) dt +h.c.
\end{align}
This corresponds to minimizing the molecular energy at the final time $T$, under the constraint that the wavefunction is obtained as a proper solution to the device Schr\"odinger equation. Here,  $\bra{\lambda(t)}$ and $\ket{\lambda(t)}$ are treated as Lagrange multipliers. 
The control function \footnote{conventionally known as control Hamiltonian but we use the edited name to avoid confusion with drive Hamiltonian ($\bar{H}_C$)} may be defined following Pontryagin's maximum principle~\cite{liberzon2011calculus,lin2019application,brady2021optimal} and Eq.~\eqref{eq1}, taking the form 
\begin{align}
    \mathbb{H}&=\langle \lambda(t)|i\hat{H}_{I,C}(t)|\Psi(t)\rangle-\langle \Psi(t)|i\hat{H}_{I,C}(t)|\lambda(t)\rangle,
\end{align}
with the time evolution of state $|\Psi(t)\rangle$ and co-state  $|\lambda(t)\rangle$, and end point of co-state $|\lambda(T)\rangle$ given by
\begin{align}
    & |\dot\Psi(t)\rangle=-iH_{I,C} |\Psi(t)\rangle , \label{eq4a}\\
    &|\dot \lambda(t)\rangle=-iH_{I,C} \lambda(t),\label{eq4b}\\
    &|\lambda(T)\rangle=\hat{H}_{\mol}|\Psi(T)\rangle. \label{eq4c}
\end{align}
The derivation of the control function $\mathbb{H}$ and details of the application of Pontryagin's principle are given in the SI. 
At the optimal control ($\Omega^*(t)$,$\Psi^*(t)$,$\lambda^*(t)$ where $^*$ denotes the optimal path), Pontryagin's maximum principle states that
\begin{align}\label{ponts1}
\mathbb{H}(t,\Omega_n^*(t), \Psi^*(t), \lambda^*(t))\geq \mathbb{H}(t,\Omega_n(t), \Psi^*(t),\lambda^*(t)),
\end{align}
at each point in time.
Notice that the $\mathbb{H}$ formed in ctrl-VQE is an explicit function of time.  $\mathbb{H}$ may be written as a sum of $\mathbb{H}$ for each qubit as 
\begin{align}
    \mathbb{H}=\sum_q \mathbb{H}_q,
\end{align}
where $q$ is the qubit index. For ctrl-VQE, the control function for each qubit may be written as
\begin{align}
    \begin{split}
        \mathbb{H}_q=&2\text{Re}\langle \lambda(t)|-i\Omega_q(t) e^{i\hat{H}_Dt}(e^{i\nu_qt} a_q\\
        &+e^{-i\nu_qt} a^{\dagger}_q)e^{-i\hat{H}_Dt}|\Psi(t)\rangle,\\
        =&\Omega_q(t)\phi_q(t),
    \end{split}
\end{align}
where the switching function $\phi_q(t)$ is given by
\begin{align}
    \begin{split}\label{eq5b}
    \phi_q(t)=&2\text{Re}\langle \lambda(t)|-i e^{i\hat{H}_Dt}(e^{i\nu_qt} a_q+e^{-i\nu_qt} a^{\dagger}_q)\\&e^{-i\hat{H}_Dt}|\Psi(t)\rangle.
\end{split}
\end{align}
Through Eq.~\eqref{ponts1} to \eqref{eq5b}, it can be shown that the optimal controls $\Omega^*_q(t)$  obey the following condition,
\begin{equation}\label{pmpr}
\Omega^*_q(t) = \begin{cases}
\text{max bound} &\text{if } \phi_q(t)>0\\
\text{min bound} &\text{if } \phi_q(t)<0\\
\text{undefined} &\text{if } \phi_q(t)=0.
\end{cases}
\end{equation}
The result of this is that  optimal controls follow a ``bang-bang'' structure when the switching function has a non-zero value; specifically the controls are maximized at time $t$ in the case of positive $\phi_q(t)$ and vice versa.
The optimal solution is undefined in the case of a singular switching function $\phi_q(t)=0$ as the control function becomes independent of control parameters in that case.
This condition, $\phi_q(t)=0$, is well studied and is known as the ``singularity'' condition \cite{liberzon2011calculus}. 
For a single-qubit system, Eq. \ref{eq5b} is zero for an extended period of time only when $\lambda(t)=\Psi(t)$ for an extended period. This is not possible in the electronic structure problem due to Eq.~\eqref{eq4c}; therefore, the optimal protocol will follow a bang-bang form for a single qubit. To study the case of two qubits, as in the case of the H$_2$ molecular calculation, we investigate the optimal pulse shape by carrying out ctrl-VQE transmon simulations and limiting the pulse duration to the requisite MET. 

In agreement with the analytic results, our simulations reveal the emergence of bang-bang pulse controls for time-optimal simulations of the H$_2$ ground state at a bond length of $1.5~\text{\AA}$.
Fig. \ref{fig2} presents the pulse shapes obtained after optimization at different pulse durations for both 2-level qubits (Fig. \ref{fig2}(a)) and 3-level qutrits (Fig. \ref{fig2}(b)). 
Similar to the cartoon depiction in Fig. \ref{fig1}, there are many ways of reaching the final state when the evolution time exceeds the MET (i.e., at $T>T^*$).
In the top two panels, we present one of the many solutions found, 
noting that both qubits and qutrits exhibit a highly unstructured pulse shape. 
In the middle two panels ($16$~ns and $10$~ns, respectively), 
both qubits and qutrits begin to develop a bang-bang-like structure. 
As the pulse duration is further reduced, the pulse shape becomes increasingly more uniform until it reaches a well organized bang-bang shape near a MET of $15$~ns and $8.94$~ns for qubits and qutrits, respectively. 

\begin{figure}[t]
    \centering
    \subfloat[qubit 1 \label{fig4a}]{
    \includegraphics[width = 0.45 \textwidth]{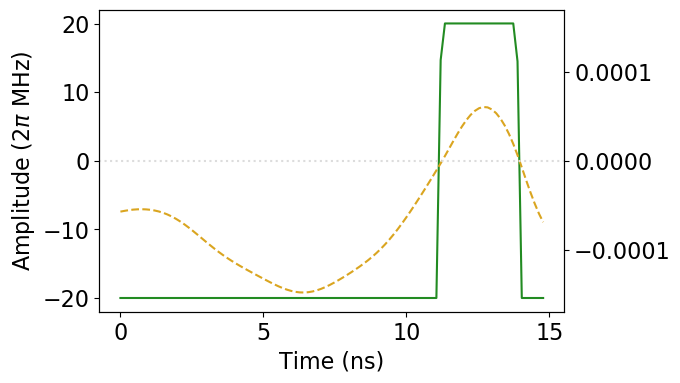}
    }\\
\subfloat[qubit 2 \label{fig4b}]{
    \includegraphics[width= 0.45 \textwidth]{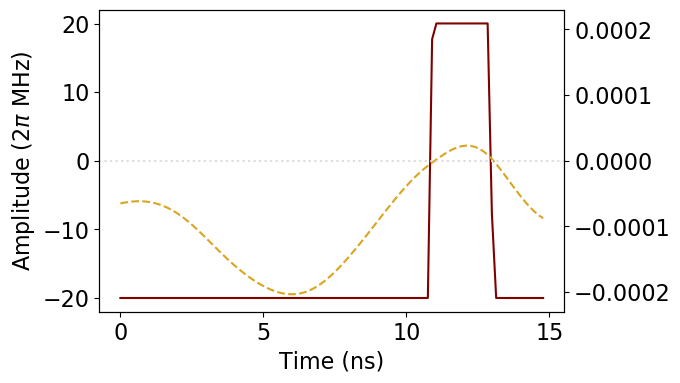}
}
\caption{(a) The switching function $\phi_1(t)$ (yellow dashed) and pulse shape $\Omega_1^*(t)$ (green) for the optimal drive on qubit 1 in two-level transmon qubits. (b) The switching function $\phi_2(t)$ (yellow dashed) and pulse shape $\Omega_2^*(t)$ (red) for the optimal drive on qubit 2 in two-level transmon qubits. The plots are made using 100 pulse segments, 1000 Trotter steps in the simulation, and a pulse duration of $T=14.93$ ns, reaching an accuracy of $10^{-6}$ Hartree relative to the FCI energy.}
    \label{fig4}
\end{figure} 

As the evolution time is decreased (such that we reach our VQE trial state faster), the computational task of optimizing a pulse becomes increasingly difficult due to the quickly rising number of local parameter traps. 
This makes it difficult to confirm that one has actually converged to $T^*$. 
In order to verify that we are near $T^*$, in Fig. \ref{fig4}, we  compare the switching function from Eq.~\eqref{eq5b}, $\phi_q(t)$, with the optimized pulses to check the agreement with results obtained using Pontryagin's maximum principle. 
By plotting the function $\phi_q(t)$~\footnote{Here we use 100 pulse segments and 1000 Trotter steps. We tested smaller and larger Trotter steps and found no appreciable change in results.} atop the optimized pulses in Fig. \ref{fig4},
we can readily see that the pulse switching times coincide with the roots of the switching function, and that at each time, the pulse is maximized in the direction determined by the sign of $\phi_q(t)$, in agreement with Pontryagin's maximum principle,
i.e., when $\phi_q(t)$ is positive (negative), $\Omega_q(t)$ should assume the largest positive (negative) value allowed by the constraints.

For the calculation of $\phi_q(t)$, we first carry out a ctrl-VQE simulation using time-optimal controls to evaluate $\ket{\Psi(t)}$ at each time step. We then form $\ket{\lambda(T)}$, and back-propagate it using Eqs.~\eqref{eq4b} and \eqref{eq4c}. Finally, we evaluate $\phi_q(t)$ through Eq. \eqref{eq5b}. It can be observed that there is no region in time where the function $\phi_q(t)$ is 0 for a continuous period of time. Further, the times at which $\phi_q(t)$ vanishes coincide with the times at which the pulse changes sign, consistent with the analytical prediction of Eq. \eqref{pmpr}.  This can also be understood from the fact that $\phi_q(t)$ is proportional to the analytical derivative derived in Ref.~\cite{Meitei2020}. 
\subsection{Minimum evolution time for a system of transmon qudits}\label{s32}
\subsubsection{Leakage speeds up evolution}

The ctrl-VQE optimized pulses for H$_2$ shown in Fig.~\ref{fig2} demonstrate that the ground state is obtained significantly faster when higher transmon levels are available. 
The MET found for a system of qubits for this problem was 15.00 ns while the MET for a system of qutrits was 8.94 ns. This amounts to a $\sim40$\% reduction in evolution time when a qutrit system is used. 
Considering that short coherence times are one of the most restrictive bottlenecks on NISQ devices, 
this reduction in state preparation time constitutes a significant practical advantage of this approach. 

Of course, this decreased evolution time is not entirely free of cost. 
By populating higher energy states (as depicted in Fig. \ref{fig1}), the final molecular energy can only be computed by first projecting onto the computational space and renormalizing the expectation value. 
However, assuming one can resolve the difference between computational states and leakage states in a measurement, 
one can simply post-select, discarding measurements of leakage states, but keeping track of how often such measurement outcomes are obtained. 
The percentage of leakage state measurements then directly informs one how to normalize the objective function. 
With increasing weight in the leakage space, the number of discarded measurements increases, which requires one to perform an increasing number of shots.
However, this increased shot-count overhead is directly determined by the extent of the leakage, which is also partially controlled by the amplitude constraints on the pulse. 

\begin{figure}[t]
  \centering
  \includegraphics[width=0.9\linewidth]{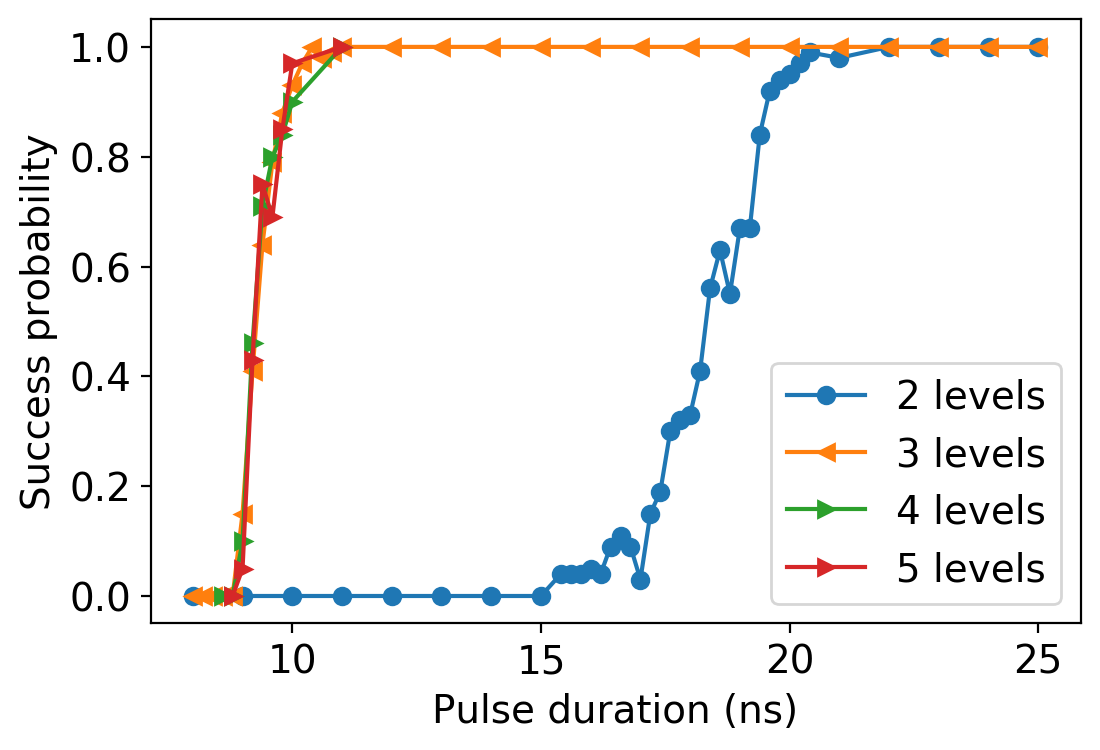}
\caption{Success probability of the simulation to reach the target state vs pulse duration. The probability is computed using 100 random initializations and with an energy error of less than $10^{-8}$ Hartree compared to the FCI solution. The plot shows that transmons truncated to two levels require more evolution time as compared to transmons with more than two levels. 
    }
    \label{fig3}
\end{figure}
In addition to the decreased time of evolution, 
access to leakage states also improves the ability to optimize the pulses themselves.
Fig. \ref{fig3} shows the probability of finding the target solution plotted against pulse duration for transmons with 2, 3, 4, and 5 levels retained. The probability is determined by using 100 random initializations and counting how many of them succeed in preparing the ground state of the $H_2$ molecule (to within $10^{-8}$ Hartree).
The results show that the probability to reach the target state drops abruptly at $\sim9$ ns for qutrits and at $\sim$19 ns for qubits.  
It is evident that the computational difficulty to reach the target state in the case of qubits is significantly higher, and it takes more optimization steps for the simulation to converge in the case of qubits as compared to qutrits (increase in optimisation steps near MET was also observed in Ref.~\cite{tibbetts2012exploring}).
This indicates that a computational phase transition occurs at the MET. 
The speedup obtained using qutrits is not further enhanced by using more than three levels in our two-transmon calculations. 
This is likely due to the fact that the higher levels are far detuned from the microwave drive, and are thus not easily populated by the microwave drive. Therefore, the contribution of these states to preparing the final state is negligible compared to that of the third transmon level.

We find that the leakage at pulse durations close to the MET for qutrits is $\sim50\%$, 
indicating a necessary 2-fold increase in the number of shots required. 
We have further seen that this leakage can be optimized to a desired fraction of the total population by introducing a penalty term in our cost function to penalize leakage over the desired amount. For instance, using a 0.01 Hartree penalty per 1\% leakage for any amount of leakage over $10\%$ of the total population (such that a leakage of 11\% will result in an energy penalty of 0.01 Hartree, while any amount of leakage below 10\% will have no energy penalty) reduces the MET from 15~ns to about 12.5~ns (see SI for details). So a penalty can be introduced in the cost function to keep the leakage below a target threshold, while still taking advantage of faster state evolution. However, we find that a penalty term that reduces the leakage to zero does not result in any speedup in our simulations. The no-leakage case is the same as performing the simulation without projecting onto the computational space (the space spanned by the $|0\rangle$ and $|1\rangle$ levels of the qudits), which has also not resulted in a faster state evolution in our simulations. This indicates that some population in the higher-lying qudit levels is necessary at the final time to achieve this faster state evolution. Another important observation is that most of the leakage lies in specific states, namely $|02\rangle$ and $|20\rangle$ in the present simulations. 

It is also worth noting that for pulse durations $T$ significantly greater than the minimum time $T^*$, we do not observe any traps (getting stuck in local minima) for either qubits or qutrits. This can be understood through results in quantum optimal control which state that a controllable system is free from traps unless the drive is severely constrained \cite{tibbetts2012exploring}. Once the drive has a duration significantly greater than the minimum duration needed for the state evolution, the system reaches the target state without getting hindered by traps. Local traps are often encountered in gate-based VQEs unless a minimum number of parameters are included in the circuit (over-parameterization). A major benefit of the reduced MET is that it allows this trap-free (analogous to over-parameterization in gate-based VQE) region to be reached with shorter pulses in ctrl-VQE.

\subsubsection{Mechanism of evolution speedup in qudits}
\begin{figure}[t]
\centering
\subfloat[Qubits\label{fig5a}]{
    \includegraphics[width = 0.45 \textwidth]{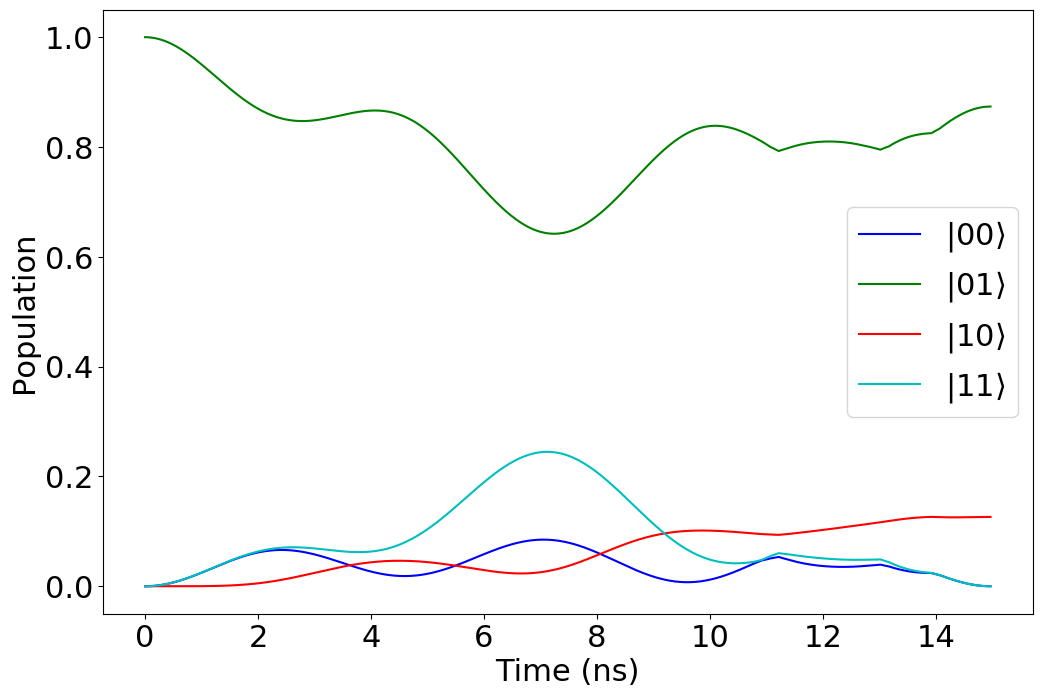}
}

\subfloat[Qutrits \label{fig5b}]{
    \includegraphics[width = 0.45 \textwidth]{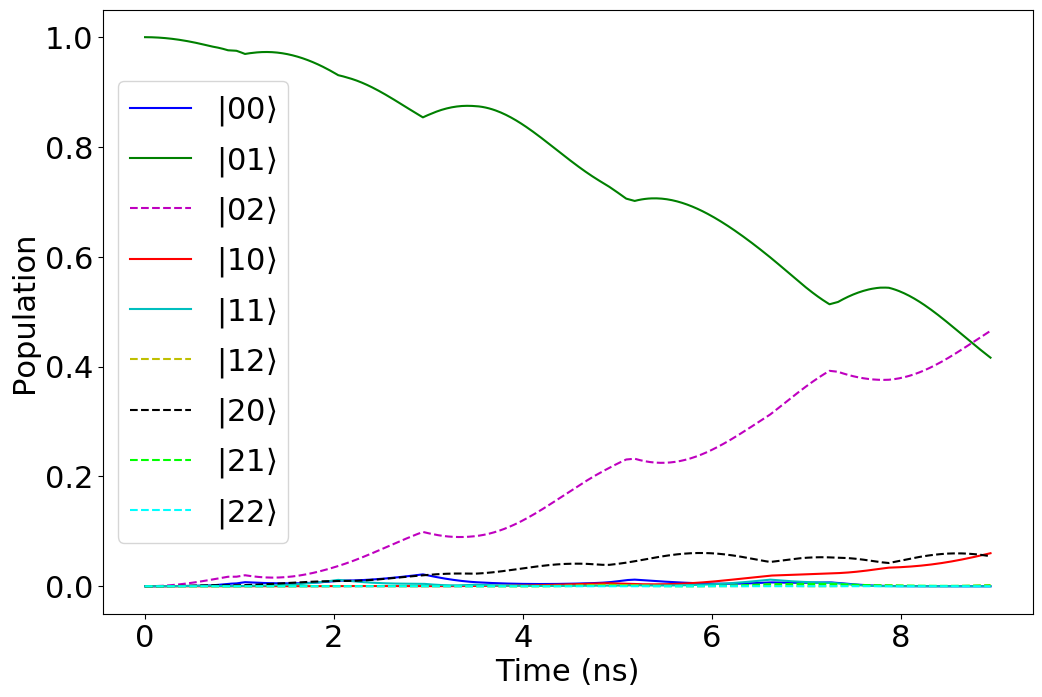}
}
%
    \caption{Population in various basis states vs evolution time for (a) qubits and (b) qutrits. The solid lines are states in the computational space, while the dashed lines are states that involve the extra transmon level. The initial state is the HF state, while the final state is the FCI wavefunction of H$_2$ at a bond distance of 1.5~\AA.}
    \label{fig5}
\end{figure} 

\begin{figure}[t]
\centering
\subfloat[Two qubits \label{fig7a}]{
\includegraphics[width = 0.45\textwidth]{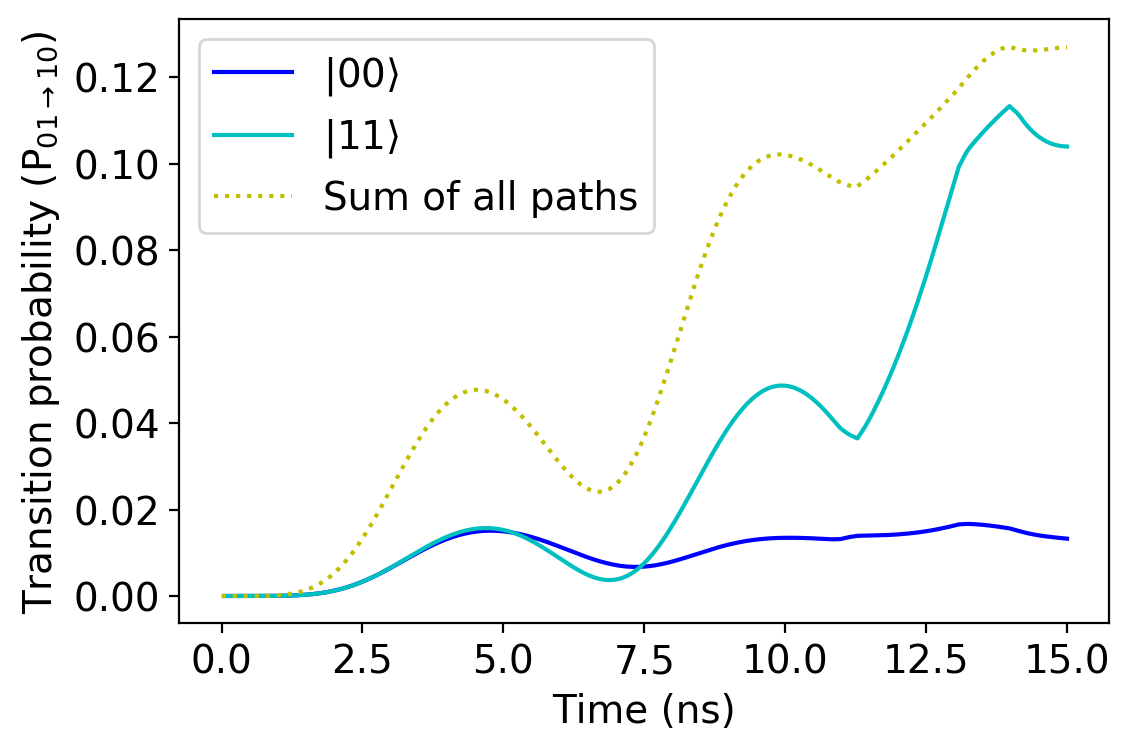}
}

\subfloat[Two qutrits \label{fig7b}]{
\includegraphics[width = 0.45 \textwidth]{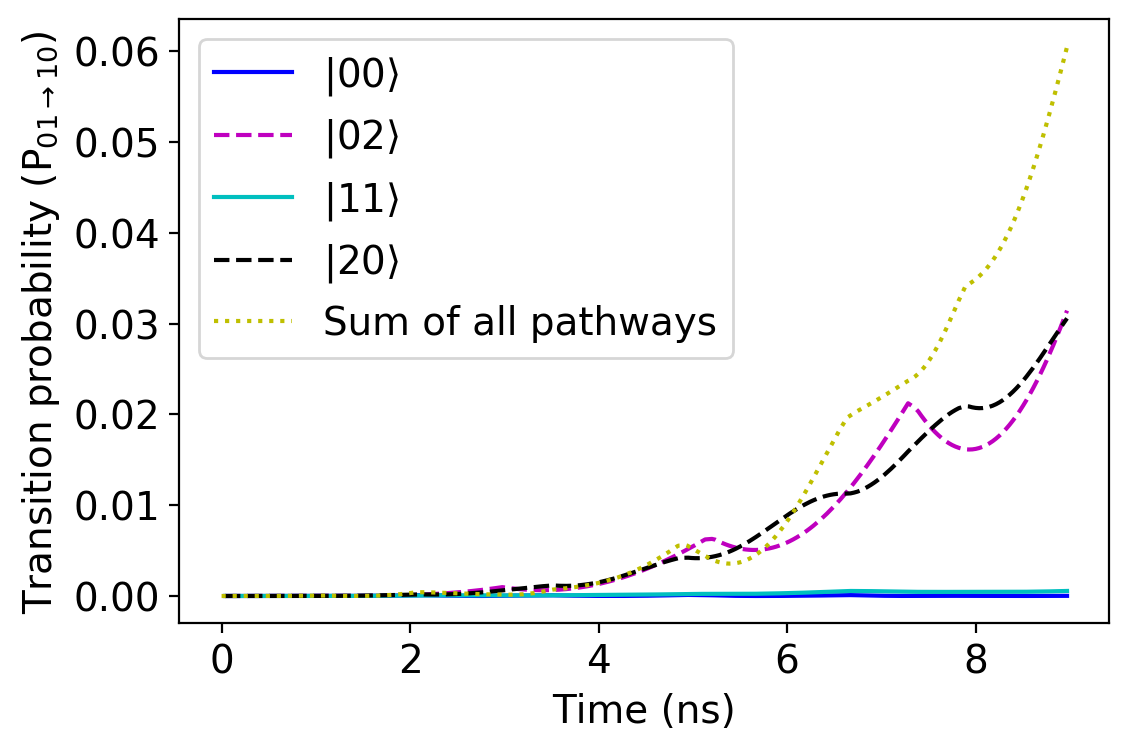}
}
%
\caption{Transition probability from state $|01\rangle$ to $|10\rangle$ through various intermediate states ($m$ in Eq. \ref{eq221}) for (a) two qubits and (b) two qutrits. The dotted yellow line shows the sum of contributions from all channels to the state $|10\rangle$.}
    \label{fig7}
\end{figure}

\begin{figure}[t]
\centering
\subfloat[Two qubits \label{fig9a}]{
\includegraphics[width = 0.23 \textwidth]{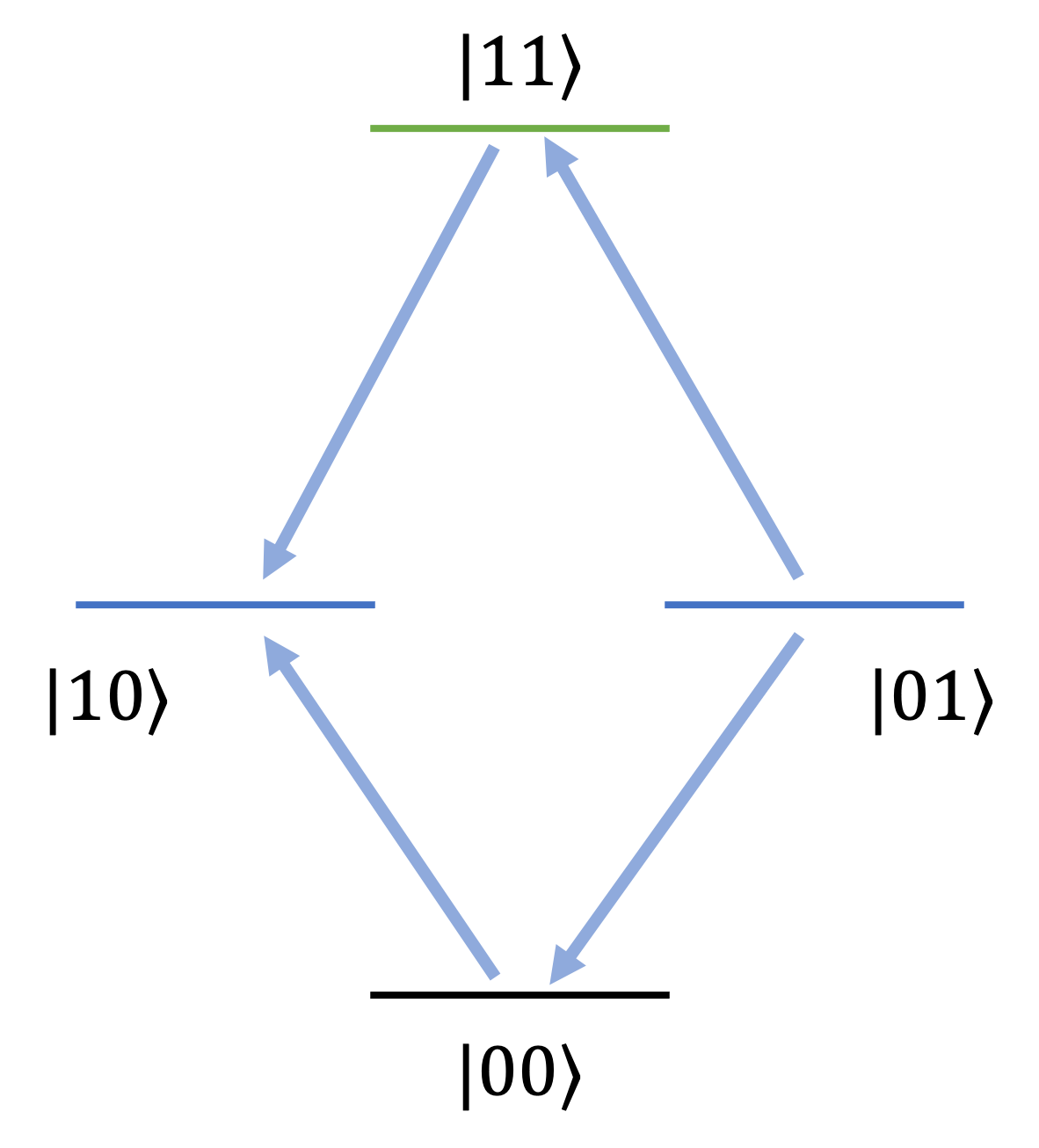}
}
\subfloat[Two qutrits \label{fig9b}]{
\includegraphics[width = 0.23 \textwidth]{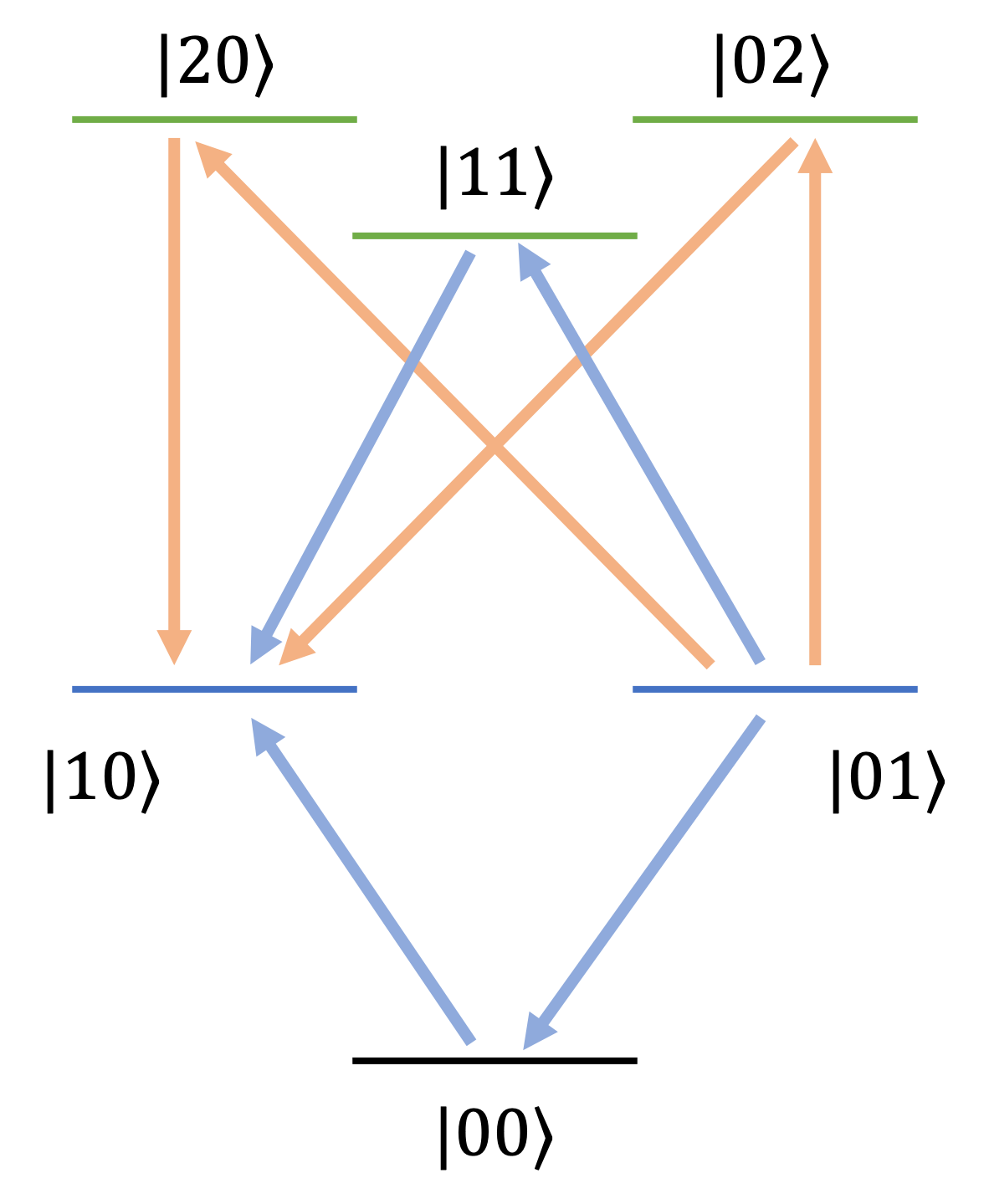}
}
    \caption{Major channels involved at second order in evolution from state $|01\rangle$ to $|10\rangle$ in the case of (a) two qubits and (b) two qutrits. The orange  channels are the new channels that arise in the qutrit case.}
    \label{fig9}
\end{figure}

We investigate the mechanism of the observed speedup using a numerical analysis based on time-dependent perturbation theory to determine which processes contribute the most toward reaching the target wavefunction in the case of qubits and qutrits. In the H$_2$ molecular problem, the initial state is the HF state $\ket{01}$. The FCI target state is a linear combination of $|01\rangle$  and $|10\rangle$, where the population in the former is $6.92$ times that of the latter. Thus, some population is transferred from $|01\rangle$ to $|10\rangle$ during the evolution. The population of each state as a function of time is shown in Fig.~\ref{fig5}, both for transmon qubits (a) and qutrits (b). It is evident that the evolution is very different between the two cases.
In the case of qubits, the states $|00\rangle$ and $|11\rangle$ get populated during the evolution and depopulate completely by the end of the evolution. These two states assist with the transfer of population from $|01\rangle$ to $|10\rangle$. In the case of qutrits, the states $|00\rangle$ and $|11\rangle$ receive very little population throughout the evolution, while states $|02\rangle$ and $|20\rangle$ exhibit a significant population buildup that remains at the end of the evolution. This striking disparity indicates that higher transmon levels provide alternate channels for reaching the target state.
To determine the exact transition channels at work, we analyze the evolution using time-dependent perturbation theory. The unitary evolution operator $U(0,T)$ that carries out the state evolution, $|\Psi(T)\rangle=U(0,T)|\Psi(0)\rangle$,
can be expanded in a Dyson series:
\begin{align}
    U(0,t_f)& = I + (-i)\int_{0}^{t_f}  dt H_{I,C}(t) \nonumber\\
    & + (-i)^2 \int_{0}^{t_f} d t_1 \int_{0}^{t_1} d t_2 H_{I,C}(t_1)H_{I,C}(t_2)+\dots
\end{align}
We consider the transition from the initial state $\ket{01}$ to the final state $\ket{10}$.
The zeroth-order transition amplitude is zero, $\langle 01 \vert 10\rangle =0$, while the first- and second-order transition amplitudes from state $|01\rangle$ to $|10\rangle$ are
\begin{align}
    A_{01\rightarrow10}^{(1)}=& \int_{0}^{t_f} \langle 10|(-i)H_{I,C}(t)|01\rangle dt,\\
    A_{01\rightarrow10}^{(2)}=& \int_{0}^{t_f} dt_1 \int_{0}^{t_1} dt_2 \langle 10|(-i)^2 H_{I,C}(t_1)H_{I,C}(t_2) |01\rangle.
\end{align}
The second-order transition amplitude can be rewritten as a sum over transitions through intermediate states: 
\begin{align}
    A_{01\rightarrow10}^{(2)}
    =&\sum_m \int_{0}^{t_f} dt_1 \int_{0}^{t_1} dt_2\langle 10|(-i) H_{I,C}(t_1)|m\rangle  \nonumber \\
    & \quad \times  \langle m|(-i)H_{I,C}(t_2)|10\rangle ,\\
    = & \sum_m A_{01\rightarrow m \rightarrow 01},
    \label{eq221}
\end{align}
where the summation is over all the eigenstates of the transmons.
We have checked that evolving the wavefunction using the second-order evolution operator $U^{(2)}(0,T)$ yields the target state up to an infidelity of $\sim 0.01$, and hence second order is sufficient for a qualitative study of the state evolution. The first-order transition probability, $A_{01\rightarrow10}^{(1)}=0$, vanishes for both qubits and qutrits, as the static coupling between transmon qubits is treated as a part of the device Hamiltonian, which is diagonal in the computational basis. Thus the transition from initial to final state is entirely through higher-order processes.

\begin{figure}[t]
\centering
\subfloat[Two qubits \label{fig8a}]{
\includegraphics[width = 0.45 \textwidth]{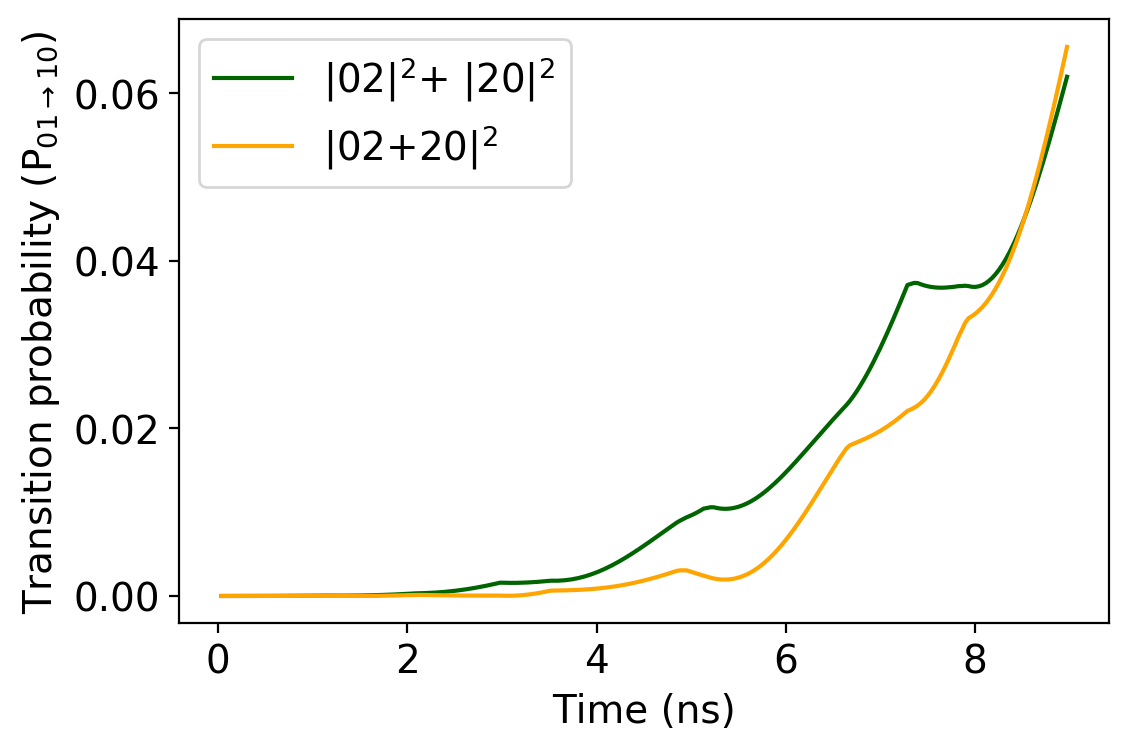}
}

\subfloat[Two qutrits \label{fig8b}]{
\includegraphics[width = 0.45 \textwidth]{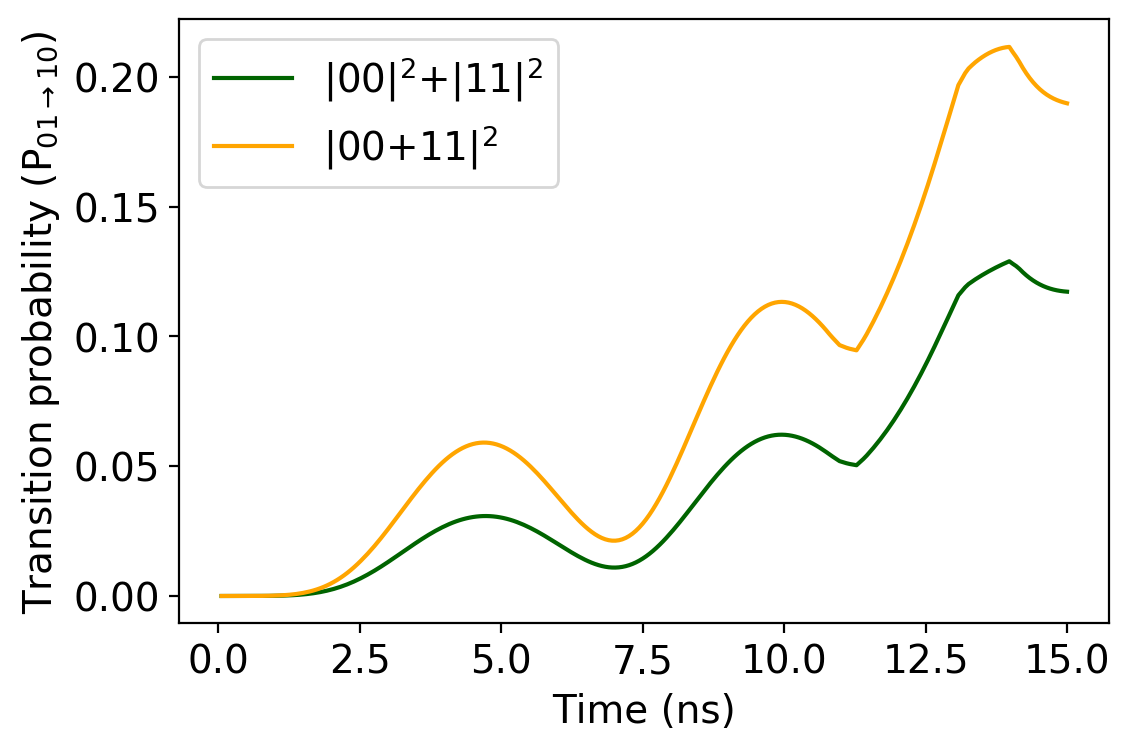}
}
%
\caption{Role of constructive interference between the main transition channels in the case of (a) two qubits (b) two qutrits. The constructive interference (difference between orange and green lines) is more significant in the qubit case.}
\label{fig6}
\end{figure} 

We show the second-order transition probability from state $\ket{01}$ to $\ket{10}$ for two transmon qubits and two transmon qutrits in Figs.~\ref{fig7}a and~b, respectively. The second-order transition probability is defined as
\begin{equation}
    P_{01 \rightarrow 10} \equiv \left\vert A_{01 \rightarrow 10}^{(2)} \right\vert^2 = \left\vert \sum_m A_{01 \rightarrow m \rightarrow 10}^{(2)} \right\vert^2.
\end{equation}
To investigate which transition path is most important in the time-optimal evolution, we compare the contributions to the transition probability from two dominant paths. In the qubit case, the paths are the ones that go through states $\ket{00}$ and $\ket{11}$ (see Fig.~\ref{fig7}a orange line) while for the qutrit case, the dominant paths are the ones through $\ket{02}$ and $\ket{20}$ (see Fig.~\ref{fig7}b orange line). Note that in both cases, the truncated transition probability is close to the second-order one, which shows that the transition is predominantly happening through those channels.
Thus, the higher-lying leakage states are participating and providing alternate channels for state evolution. This can be visualized with the help of the schematic level diagram shown in Fig.~\ref{fig9}. Basis states $|02\rangle$ and $|20\rangle$ do not exist in qubits, while this channel starts to become available in qutrits. Further, the population of basis states $|00\rangle$ and $|11\rangle$ is zero in the target state. This puts a constraint on the dynamics of qubits as it has to depopulate these basis states at the end of the evolution, while these basis states do not have any population in the case of qutrits. 
Furthermore, in Fig.~\ref{fig6}a and~b, we compare the truncated transition probability with the squares of the individual transition amplitudes. If the two transition paths constructively interfere with each other, we expect
\begin{align}
\begin{split}
    \vert A_{01\rightarrow m_1 \rightarrow 10} &+ A_{01\rightarrow m_2 \rightarrow 10} \vert ^2 \\
    &> \vert A_{01\rightarrow m_1 \rightarrow 10}
    \vert^2 + \vert A_{01\rightarrow m_2 \rightarrow 10} \vert ^2.
\end{split}
\end{align}
This is observed in the qubit case, while it is not true for the qutrit case (see Fig.~\ref{fig6}). Details for analysis using perturbation theory is presented in the SI.

\begin{figure}[t]
    \begin{center}
  \centering
  \includegraphics[width=0.9\linewidth]{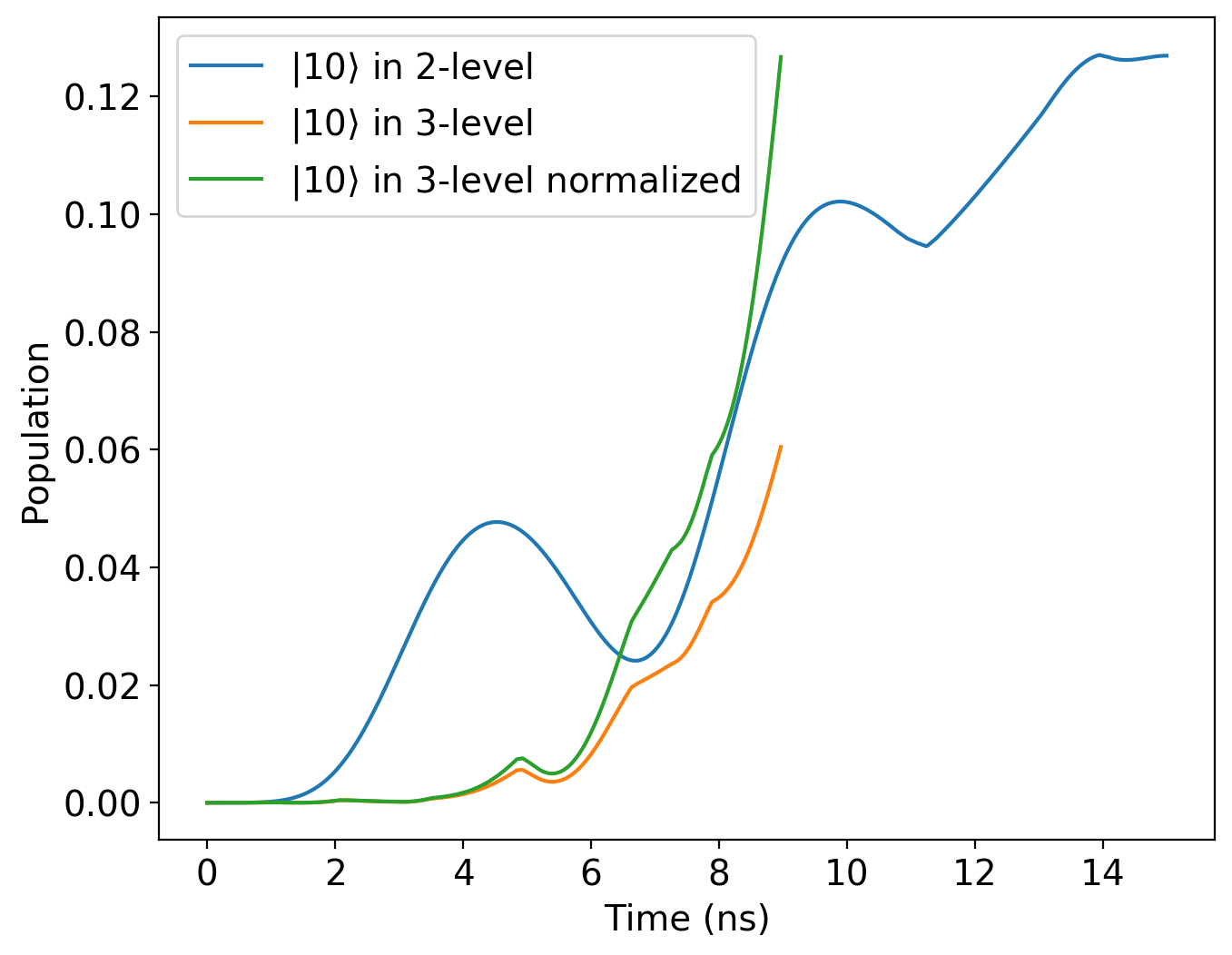}
  \centering
    \end{center}
    \caption{Population buildup in state $|10\rangle$ vs evolution time in the case of two qubits (2-level) and two qutrits (3-level). The basis state $|10\rangle$ reaches the target population faster in the case of qutrits when the final state is projected onto the computational subspace and normalized (3-level normalized).}
    \label{fig10}
\end{figure}

Finally, an additional contribution to the speedup evident in the qutrit case comes from the classical step of projecting the final qutrit state onto the computational subspace and normalizing it. This can be seen from Fig.~\ref{fig10}, where the population in state $|10\rangle$ is plotted as a function of the evolution time in the case of qubits and qutrits (unnormalized and normalized cases). It can be seen that the population in qutrits increases faster than qubits and reaches the target population faster when the final state on qutrits is projected onto the computational basis and normalized. This is because the population required by the target state in the computational subspace is smaller when we finally normalize the state. The presence of additional channels for state evolution, along with reduced population demand in the desired target state in the computational subspace together leads to a faster state evolution in qutrits compared to qubits.

Although we have focused on the electronic structure problem in the case of the H$_2$ molecule ground state in our current work, the key results are expected to generalize. This is because any problem can be mapped to the task of finding trajectories that connect an initial and target state in Hilbert space. Changing the problem would correspond to changing these initial and target states. Including higher levels will generally provide new channels for state evolution that provide shorter paths to the target state.

\section{Summary}\label{s4}
In this paper, we presented time-optimal control fields that prepare target molecular ground states on transmon quantum processors. These optimal controls were obtained using the pulse-level VQE algorithm known as ctrl-VQE. We found that the optimal controls converge to a bang-bang form, i.e., they saturate pulse amplitude constraints, as the total evolution time is reduced. We showed that this behavior is consistent with analytical results derived from Pontryagin's maximum principle. These results suggest that ctrl-VQE prepares target states on quantum computers in the shortest possible time allowed by quantum speed limits. 

In addition, we investigated how the minimal state preparation time is impacted by truncating the transmons to two or more levels. We find that the inclusion of higher transmon levels can lead to a substantial reduction in the time needed to reach the target state. We showed that this speedup is the result of two effects caused by the additional levels: (i) an enlargement of the space of wavefunctions that have perfect overlap with the target state in the computational subspace, and (ii) the appearance of new transition paths that connect the initial and target states. Together, these effects cut down the evolution time by $\sim40$\% for the simulations performed in this work.

Given the limited coherence times in NISQ devices, it is crucial that we minimize the time it takes to prepare target wavefunctions. Our results indicate that ctrl-VQE does this. Our work also shows that leakage to higher transmon levels is not always problematic. It can potentially be used to reduce the coherence time requirements needed for VQEs. 
Another important benefit of faster state evolution is that the overparameterized region, where the VQE algorithms become free from local minima (trap-free), is reached faster in qudits  as compared with qubits. Local traps have been shown to prevent gate-based VQE algorithms from  reaching the solution to the desired accuracy unless a significantly large gate depth is achieved \cite{larocca2021theory}.  


\section{Acknowledgements}
This research is supported by the Department of Energy (DOE). E.B. and N.J.M acknowledge Award No.  DE-SC0019199, and S.E.E. acknowledges the DOE Office of Science, National Quantum Information Science Research Centers, Co-design Center for Quantum Advantage (C2QA), Contract No. DE-SC0012704. A.A. would like to thank Dr. Luke Bertels and Dr. Sheril Avikkal Kunhippurayil for helpful discussions. The authors thank the Advanced Research Computing (ARC) facility at Virginia Tech for the computational infrastructure.

\bibliography{paper}

\end{document}